\documentclass[5p,twocolumn,times,number]{elsarticle}
\usepackage{amsmath}
\usepackage{amssymb}
\usepackage{graphics}
\usepackage{graphicx}
\usepackage{epsfig}
\usepackage{dcolumn}
\usepackage{bm}
\usepackage{lineno}

\begin{document}
\begin{frontmatter}

\title{A simple technique for gamma ray and cosmic ray spectroscopy using plastic scintillator}
\author[label1]{Akhilesh~P.~Nandan}
\author[label2]{Sharmili~Rudra}
\ead{sr.phys@gmail.com}
\author[label3]{Himangshu~Neog}
\author[label3]{S.~Biswas\corref{cor}}
\ead{saikat.ino@gmail.com, s.biswas@niser.ac.in, saikat.biswas@cern.ch}
\author[label4]{S.~Mahapatra}
\author[label3]{B.~Mohanty}

\author[label4]{P.~K.~Samal}

\cortext[cor]{Corresponding author}

\address[label1]{Indian Institute of Science Education and Research, Computer Science Building, College of Engineering Trivandrum Campus, Trivandrum-695 016 Kerala,India}
\address[label2]{Department of Applied Physics, University of Calcutta, 92, APC Road, Kolkata - 700 009, West Bengal, India}
\address[label3]{School of Physical Sciences, National Institute of Science Education and Research, Bhubaneswar - 751005, Odisha, India}
\address[label4]{Physics Department, Utkal University, Vani Vihar, Bhuvaneswar - 751 004 , India}

\begin{abstract}
A new and simple technique has been developed using plastic scintillator detectors for gamma ray and cosmic ray spectroscopy without single channel analyzer (SCA) or multichannel analyzer (MCA). In these experiments only a leading edge discriminator (LED) and NIM scalers have been used. Energy calibration of gamma spectra in plastic scintillators has been done using Co$^{60}$ and Cs$^{137}$ sources. The details experimental technique, analysis procedure and experimental results has been presented in this article.   
\end{abstract}
\end{frontmatter}

\section{Introduction}
Special type of plastic scintillator can be used for gamma ray spectroscopy \cite{BLR12}. Normal plastic scintillator detector, used for triggering particle in high energy physics experiments can also be used for gamma ray spectroscopy. A simple and new technique has been introduced with plastic scintillator paddle for gamma ray and cosmic ray spectroscopy without using single channel analyzer (SCA) or multi channel analyzer (MCA). In this technique only a Leading Edge Discriminator (LED) and scaler have been used. Using this technique $\gamma$~-~ray spectrum has been obtained for Co$^{60}$ and Cs$^{137}$ sources. Cosmic ray muon pulse height spectrum i.e. the spectrum for minimum ionizing particle (MIP) has also been obtained. The experimental technique and the data analysis technique have been described in Section~\ref{ex_tech} and Section~\ref{ana_tech} respectively, the experimental results have been described in Section~\ref{result}. Finally in Section~\ref{con} we have presented the conclusion of the study and a brief future outlook on the work.

\section{Experimental technique}\label{ex_tech}

A new technique has been developed for the gamma ray spectrum and cosmic ray pulse height spectrum without using single channel analyzer (SCA) or multi channel analyzer (MCA). Only a leading edge discriminator and scaler have been used. In the measurements two plastic scintillator paddles of dimension 7~cm $\times$ 7~cm and 10~cm $\times$ 10~cm with 1~cm thick plastic have been used. The scintillator detectors are named as Sc-01 and Sc-02 respectively in this article. 

For the gamma ray spectroscopy the source (Co$^{60}$ and Cs$^{137}$) has been placed on the scintillator material. Constant voltage of -~1500~V has been applied to the photo multiplier tube (PMT) of Sc-01 for signal collection. In a separate study we have observed that the efficiency of the detectors reaches a constant (plateau) value close to 100\% at high voltage beyond -~1450~V. The signals from the scintillator have been fed to the discriminator. The discriminator threshold has been increased in equal interval of 0.5~mV from an initial threshold value of 7.5~mV. For each discriminator threshold setting the count rate has been measured with the source and also without source, taking the total counts in the scaler in a particular time. 

For the cosmic ray muon detection a constant voltage of -~1500~V has been applied to the photo multiplier tube (PMT) attached to the Scintillator Sc-01 but the discriminator threshold has been increased in equal interval of 2~mV from an initial threshold value of 12~mV. In this case no source is used and the data is analyzed in a different way (as explained in Section~\ref{ana_tech}) the measurent has been started from an initial threshold value of 12~mV as lower than this value the singles count rate was very high due to noise effect. The interval is taken 2~mV as for the cosmic ray spectrum 0.5~mV is very small to get a reasonable difference betwwn two consecutive values. In this case also the count rate has been measured for each discriminator threshold settings. Data for each discriminator threshold settings has been taken for 30 minutes or more for the cosmic ray.

\section{Data analysis technique}\label{ana_tech} 

All the data analysis has been performed using ROOT, a data analysis framework developed by CERN \cite{ROOT}. In the case of gamma ray spectrum the difference of the count rate with and without the gamma ray source for a particular threshold setting gives the gamma ray count rate (say C$_{1}$) at that threshold. Likewise the gamma ray count rate (say C$_{2}$) has been measured for the next threshold setting. The difference of gamma count rate for two consecutive threshold settings gives the gamma ray signal count rate (say C=C$_{1}$-C$_{2}$) with pulse height between those consecutive threshold values. The average of these two threshold values have been taken and the C has been said to be gamma ray count rate of pulse height at that average value. The count rate has been plotted as a function of the discriminator threshold i.e. as a function of the pulse height (shown in Fig.~\ref{co60} and Fig.~\ref{cs137} and discussed in detail later).

The cosmic ray muon pulse height spectrum has been obtained in the same way as described above. In this case also the difference of count rates for two consecutive threshold settings have been calculated (say C=C$_{1}$-C$_{2}$) and assigned to pulse height with the average of these two threshold values. Here only one assumption has been taken that the noise level are same for the two consecutive threshold settings, which is not true at the lower threshold value. This is the reason of a noise peak in the MIP spectra al low pulse height region, which will be described in detail in the Section~\ref{result}.

\section{Result}\label{result} 



\begin{figure}[htb!]
\begin{center}
\includegraphics[scale=0.4]{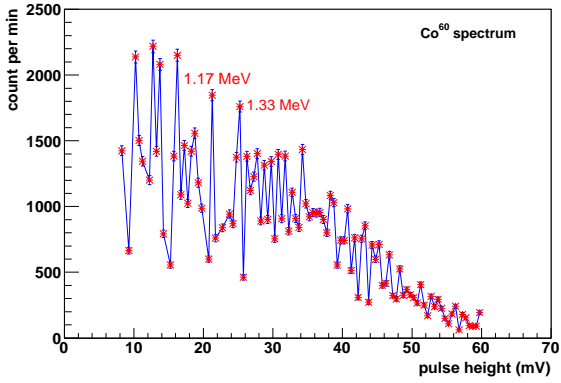}
\caption{\label{co60}Co$^{60}$ spectrum, a curve of count per minute as a function of pulse height.} \label{co60}
\end{center}
\end{figure}

\begin{figure}[htb!]
\begin{center}
\includegraphics[scale=0.4]{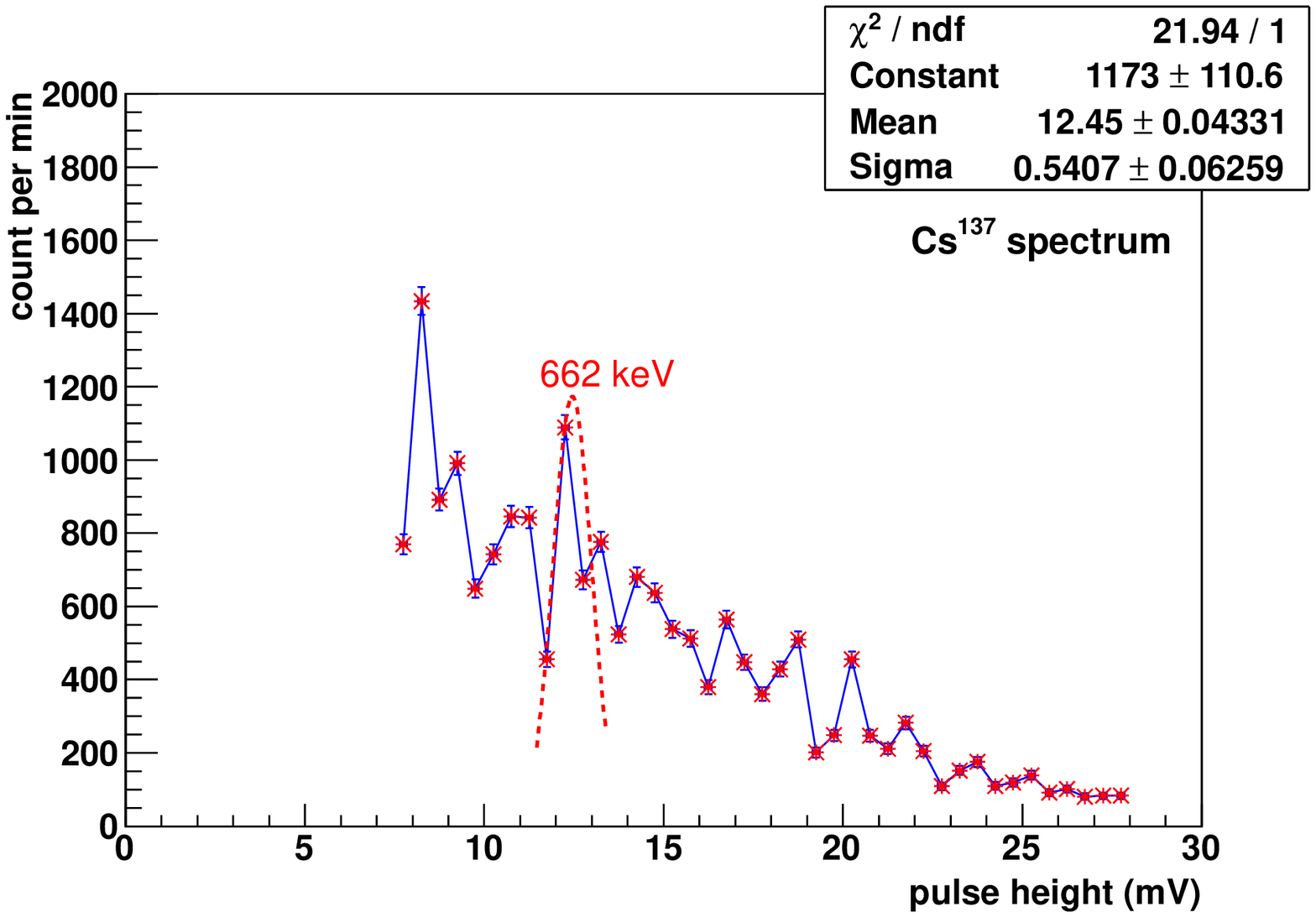}
\caption{\label{cs137}Cs$^{137}$ spectrum, a curve of count per minute as a function of pulse height.}\label{cs137}
\end{center}
\end{figure}
\begin{figure}[htb!]
\begin{center}
\includegraphics[scale=0.4]{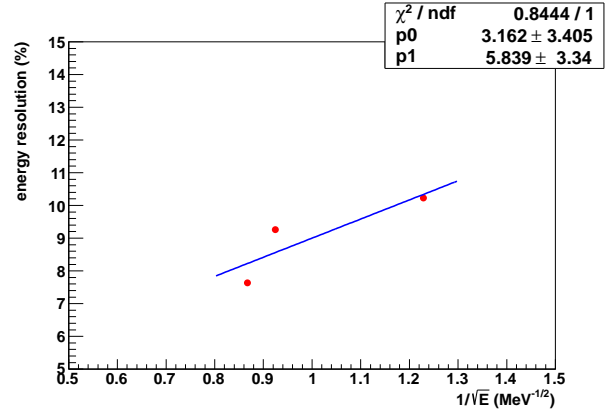}
\caption{\label{reso}Energy resolution as a function of $\frac {1}{\sqrt E}$.} \label{reso}
\end{center}
\end{figure}
\begin{figure}[htb!]
\begin{center}
\includegraphics[scale=0.4]{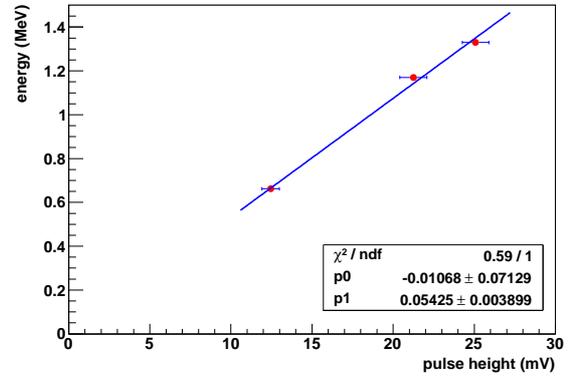}
\caption{\label{calib}Calibration curve, the linear relationship between the energy and the experimentally measured pulse height.} \label{calib}
\end{center}
\end{figure}
For the gamma ray spectroscopy Co$^{60}$ and Cs$^{137}$ sources have been used. Gamma ray spectrum for Co$^{60}$ source has been plotted in Figure~\ref{co60}. The characteristic energy for Co$^{60}$ gamma source, 1.17~MeV and 1.33~MeV have been identified and the corresponding pulse heights have been measured fitting those peaks with Gaussian function. The corresponding pulse heights have been found to be 21.25~mV and 25.08~mV respectively. The energy resolution for the detector in \% defined as, ${\sigma \times 2.355\over pulse~height}{\times 100}$ \%, has been found to be 9.3\% and 7.6\% for the 1.17~MeV and 1.33~MeV peak respectively. Here $\sigma$s and the pulse heights have been found by fitting the 1.17~MeV and 1.33~MeV gamma ray peaks for Co$^{60}$ source by Gaussian function. For Gaussian function the $\sigma$ and full width at half maxima (FWHM) are related by FWHM = $\sigma$ $\times$ 2.355. Gamma ray spectrum for the Cs$^{137}$ source has been obtained by the same method as described above and plotted in Figure~\ref{cs137}. The characteristic 662~keV peak has been fitted with Gaussian curve and the peak pulse height has been determined. The peak pulse height for the Cs$^{137}$ gamma source has been found to be 12.45~mV. In this case the energy resolution value has been found to be 10.2\%. Energy resolution obtained for the two peaks of Co$^{60}$ and one peak of Cs$^{137}$, as a function of $\frac {1}{\sqrt E}$ is plotted in Figure~\ref{reso}. The points are fitted with a straight line. An energy calibration curve has been drawn for Sc-01 at -~1500~V operating voltage, taking two points obtained from Co$^{60}$ spectrum and one point obtained from Cs$^{137}$ spectrum \cite{ERS08}. The calibration curve has been fitted with a straight line. The calibration curve has been shown in Figure~\ref{calib}. From the fit parameters it is clear that the curve nearly passes through the origin and the calibration factor is 0.054~MeV/mV. The linear relationship between the energy and the experimentally measured pulse height has been observed.

\begin{figure}[htb!]
\begin{center}
\includegraphics[scale=0.4]{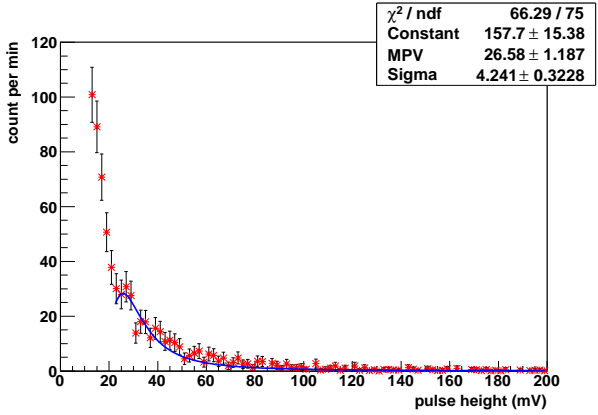}
\caption{\label{mip1}Cosmic ray (MIP) spectrum fitted with Landau function from detector Sc-01.}\label{mip1}
\end{center}
\end{figure}
\begin{figure}[htb!]
\begin{center}
\includegraphics[scale=0.4]{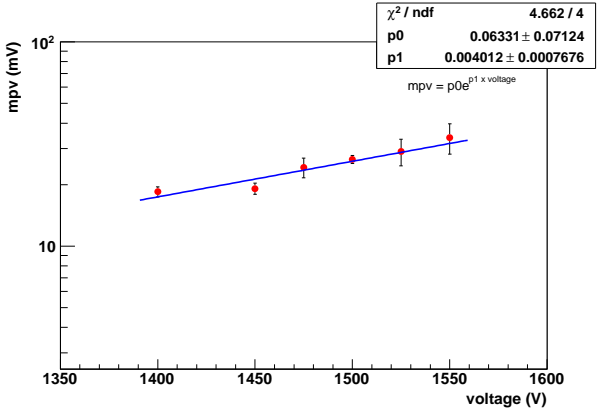}
\caption{\label{mpv1}MPV as a function of the applied voltage detector Sc-01.}\label{mpv1}
\end{center}
\end{figure}

Cosmic ray muon pulse height spectrum has also been obtained in this set-up as described in Section~\ref{ex_tech} and Section~\ref{ana_tech}. The The cosmic ray induced pulse height spectra have been obtained for two detectors Sc-01 and Sc-02 and is shown in Figure~\ref{mip1} for Sc-01. Same type of spectrum has also been observed for Sc-02.  The plot has been fitted with Landau distribution; as for the cosmic ray there is a large fluctuation of energy deposition in the detector material as revealed from the $\chi^2$ per degrees of freedom. This plot has been obtained keeping the PMT voltage constant at -~1500~V. MIP peak seems to have developed around $\sim$~30~mV but large noise peak (below 20~mV) masks it. This peak come due to our assumption that the noise level are nearly same for the two consecutive threshold settings, which is not true for the very low threshold. From the cosmic ray spectra obtained from Sc-01 and Sc-02 it is also clear that at most probable value (MPV) of the count rate is nearly double for the Sc-02 relative to Sc-01 as the area of the Sc-02 (area 100 cm$^2$) which is nearly double of that of the Sc-01 (area 49 cm$^2$). As the calibration curve has been drawn for the Sc-01 at -~1500~V and the Sc-01 has been operated at -~1500~V during the cosmic ray study it is seen from Figure~\ref{mip1} that the most probable energy deposition in 1~cm thick plastic scintillator is $\sim$~1.4~MeV. The cosmic ray spectra for these two scintillators have also been obtained varying the applied voltage and the MPVs are plotted as a function of the applied voltage in Figure~\ref{mpv1}  for Sc-01. Similar plot has also been obtained for Sc-02. Both the plots are fitted with the exponential function MPV = $p_{0}e^{p_{1}.voltage}$, where the parameters are $p_{0}$ and $p_{1}$. $p_{1}$ for both the scintillators have been found to be $\sim$~0.004$\pm$0.0008. 

\section{Conclusions and outlook}\label{con} 
A simple and new technique has been developed for gamma ray and cosmic ray muon pulse height spectroscopy without using SCA or MCA. Only scintillator detector, a leading edge discriminator and a NIM scaler have been used in this technique. Gamma ray spectrum has been obtained for Co$^{60}$ and Cs$^{137}$ sources. Proportionality in energy and pulse height has been observed. The energy resolution for the detector  has been found to be 9.3\% and 7.6\% for the Co$^{60}$ 1.17~MeV and 1.33~MeV peak respectively and 10.2\% for 662~keV peak of Cs$^{137}$. Cosmic ray muon pulse height spectrum has been obtained for two scintillators and fitted with Landau distribution. The most probable energy deposition in 1~cm thick plastic material has been found to be $\sim$ 1.4~MeV. Although the energy resolution is not so good but still using plastic scintillator detector gamma spectroscopy and cosmic ray muon pulse height spectroscopy can be done. Main drawback of this technique is that this process is time consumable and may not be useful for real experiment; however, this process is very useful and can be applied for laboratory measurement where MCA or SCA are not available. 

\vspace{1 cm}

\section{Acknowledgements}
We acknowledge Mr. Subasha Rout of Utkal University,  Dr. Sudakshina Prusty, Dr. Gunda Santosh Babu, Dr. Lokesh Kumar and Mr. Rudranarayan Mohanty of NISER for helping during this study. S. Biswas acknowledges the support of DST-SERB Ramanujan Fellowship (D.O. No. SR/S2/RJN-02/2012). XII$^{th}$ Plan DAE project Experimental High Energy Physics at NISER-ALICE is also acknowledged.

\end{document}